\documentclass[12pt]{article}

\usepackage{graphicx}
\usepackage{times}
\usepackage{amsmath,amssymb}
\usepackage{enumitem}
\usepackage{color}
\usepackage{lipsum}

\let\oldabstract\abstract
\let\oldendabstract\endabstract
\makeatletter
\renewenvironment{abstract}
{%
{\list{}{\addtolength{\leftmargin}{1em} 
\listparindent 1.5em%
\itemindent    \listparindent%
\rightmargin   \leftmargin%
\parsep        \z@ \@plus\p@}%
\item\relax}%
{\endlist}%
\oldabstract}
{\oldendabstract}
\makeatother

\begin{document}
	\vskip 24pt
\centerline{\LARGE Invention as a Combinatorial Process:}
\vskip 12pt
\centerline{\LARGE Evidence from U.S. Patents}
\vskip 22pt
\centerline {\large Hyejin Youn$^{1, 2, 3, 4\ast}$, Lu\'is M. A. Bettencourt$^{4}$, Deborah Strumsky$^{5}$ and Jos\'e Lobo$^{6}$}
\vskip 22pt
\centerline{\it $^{1}$ Institute for New Economic Thinking at the Oxford Martin School, Oxford, OX2 6ED, UK. }
\vskip 4pt
\centerline{\it $^{2}$ Mathematical Institute, University of Oxford, Oxford, OX2 6GG, UK. }
\vskip 4pt
\centerline{\it $^{3}$ CABDyN Complexity Centre, University of Oxford, Oxford, UK.}
\vskip 4pt
\centerline{\it $^{4}$ Santa Fe Institute, 1399 Hyde Park Road, Santa Fe NM 87501, USA.}
\vskip 4pt
\centerline{\it $^{5}$ Department of Geography \& Earth Sciences,}
\vskip 4pt
\centerline{\it University of North Carolina-Charlotte, Charlotte, NC, USA.}
\vskip 4pt
\centerline{\it $^{6}$ School of Sustainability, Arizona State University, Tempe, AZ, USA. }
\vskip 1cm
\date{\today} 
\baselineskip 18pt

\renewcommand{\abstractname}{}
\begin{abstract}
Invention has been commonly conceptualized as a search over a space of
combinatorial possibilities. Despite the existence of a rich literature, spanning a variety
of disciplines, elaborating on the recombinant nature of invention, we lack a formal
and quantitative characterization of the combinatorial process underpinning inventive
activity. Here we utilize U.S. patent records dating from 1790 to 2010 to formally
characterize the invention as a combinatorial process.
To do this we treat patented inventions as carriers of technologies and avail ourselves of the elaborate system of technology codes used by the U.S. Patent Office to classify the technologies responsible for an invention's novelty. We find that the combinatorial inventive process exhibits an invariant rate of ``exploitation" (refinements of existing combinations of technologies) and ``exploration" (the development of new technological combinations). This combinatorial dynamic contrasts sharply with the creation of new technological capabilities -- the building blocks to be combined -- which has significantly slowed down. We also find that notwithstanding the very reduced rate at which new technologies are introduced, the generation of novel technological combinations engenders a practically infinite space of technological configurations.
\end{abstract}
\vskip 12pt


\newpage

\section{Introduction}

A common conceptualization of invention in both the biological and socioeconomic domains sees it as an adaptive search process over a space of combinatorial possibilities \cite{Kauffman1993}. A widely shared, and related, perspective in the economics and management science literature posits the combination of new and existing technological capabilities as the principal source of technological novelty and invention \cite{Allen1997,Basalla1998,VonHippel1988,Kogut1992,Tushman1992, Weitzman1998, Auerswald2000, Fleming2001, Arthur2009}. The importance of recombination in the generation of new technologies was also recognized by an early sociological, archaeological and anthropological literature on invention \cite{ Kubler1962, Tarde1903, Ogburn1922, Kaempfffert1930, Schumpeter1934, Gilfillan1935, Usher1954, Shockley1957}. Despite the recurrence of the theme of invention as resulting from a combinatorial process, discussions about this process have been based largely on case-studies and historical analyses which do not provide much by way of a quantitative characterization of the process generating inventions. A formal study of the processes that generate inventions, and more ambitiously the development of a theory of technological change, necessitates the identification of countable units of technology. Without the means to discretize technology discussions of technological combinatorics are bound to remain more a metaphor than a model.

What is a \emph{technology}? According to Romer \cite{Romer2010} technologies are ideas about how to re-arrange matter, energy and information; for Arthur \cite{Arthur2007} technologies are means to fulfill a human need or purpose. In the present discussion we similarly define technologies as artifacts, devices, methods, and materials available to humans to accomplish specific tasks. These definitions, though capturing essential features of what technology is, do not readily enable the identification of new inventions. Distinct from a technology, an \emph{invention} integrates distinct technological functionalities. We note that technological novelty is not the same as inventive novelty. Technological novelty arises, and technological change occurs, when new technological functionalities are introduced into the existing repertoire of technologies. A new invention consists of technologies, either new or already in use, brought together in a way not previously seen. The laser, for example, presupposes the ability to construct highly reflective optical cavities, create light intensification mediums of sufficient purity, and supply light of specific wavelengths; the polymerase chain reaction requires the ability to finely control thermal cycling (which involves the use of computers), and isolate short DNA fragments (which in turn applies techniques from chemical engineering); the inkjet printer involves the ability to position extremely small bits of matter with extreme precision (an ability that has been used in a variety of applications other than printing documents); the incandescent light bulb involves the use of electricity, a heated filament, an inert gas and a glass bulb.

Here we utilize U.S. patent records so as to formally characterize the invention as a combinatorial process. To do this we treat patented inventions as ``carriers" of technologies and avail ourselves of the elaborate system of technology codes used by the U.S. Patent Office to classify the technologies responsible for an invention's novelty. The technology codes provide a rich data resource for identifying individual technological capabilities, marking the arrival of new technologies and studying the role of technological combinatorics in propelling invention. Specifically we address the questions of: How prevalent have combination and recombination been as sources of invention; how has the development of new technological functionalities (as captured by the introduction of new technology codes) fueled the combinatory process; and what is the nature of the combinatorial process. A formal description of the combinatorial inventive process makes it possible to investigate how the technology space has been searched as well as to assess the ways in which the search process generates different types of inventive novelty. By examining an empirical record spanning over 200 years of inventive activity we have been able to identify surprising regularities in the generation of inventive novelty. Our discussion concludes by outlining additional possible explorations of technology space.

\section{ Patents as Footprints of Invention }
Some inventions, namely those that are patented, leave behind a documentary trail, enabling us to study the invention processes in a systemic way. According to U.S. Patent Law, a patent can be granted to the invention or discovery of a new and useful process, machine, manufacture, or composition of matter, or to any new and useful improvement thereof. The statutory definition of a patentable invention states that it must be novel, non-obvious and useful (35 U.S.C., Chapter 10, \S\S 101-103). Inventions -- new artifacts, devices, processes, materials or compounds -- thus embody technological novelty. The United States Patent and Trademark Office (USPTO) effectively defines inventions as bundles of technological capabilities, and ``technology" is in turn defined s the ``application of science and engineering to the development of machines and procedures in order to enhance or improve human conditions, or at least to improve human efficiency in some respect''\cite{USPTO1995}.

The U.S. Patent Office grants three types of patents: utility patents, also referred to as ``a patent for invention,'' are issued for the invention of ``new and useful'' processes, machines, artifacts, or compositions of matter (this type represents over ninety percent of all patents); design patents, which are granted for the ornamental design of a functional item; and plant patents which are conferred for new varieties of plants or seeds. (The results presented here use data covering the three types of patents). Although it is the case that most patents have been granted to inventions involving machines or the transformation of one physical substance into another, business methods, computer programs and algorithms can also be patented. A patent is intended to be limited to only one invention consisting of several closely
related and indivisible (i.e. integrated) technologies that, acting together, accomplish a specified task (in patent law this is known as the ``unity of invention'' principle). In plain terms what this means is that a jet engine cannot be patented but the numerous components of the jet engine can. The ``unity of invention'' principle makes it plausible to use patented inventions as means to discretize technologies since the inventions heralded by a patent are meant to be decomposable into at most a few distinct technologies.

The USPTO is required by law to ``\dots revise and maintain the classification by subject matter of United States letters patent, and such other patents and printed publications as may be necessary or practicable, for the purpose of determining with readiness and accuracy the novelty of inventions for which applications for patent are filed.'' (35 U.S.C., Chapter 1, \S8) In order to fulfill this obligation the USPTO classifies the technologies \textit{responsible for an invention's novelty} through an elaborate system of \textit{technology codes}. Technological novelty, as revealed through patented inventions, may result from the introduction of new technologies or from the combination of existing capabilities in ways that have not been previously witnessed in the patenting record. At any given time the existing set of technology codes available to a patent examiner is essentially a description of the current set of technological capabilities. With each new patent application a patent examiner must decide which existing codes, or combination of existing codes, to use to describe the technological components of the proposed invention, or whether new codes are needed to capture the invention's novelty. The introduction of a new technology code sets in motion a retroactive reclassification of all previous patents that may have embodied the newly recognized technological capability. The Patent Office's technology codes thus constitute a set of consistent definitions of technological capabilities spanning over 200 years of inventive activity.

The legal essence of a patent is the right to exclude others from practicing the invention; the legal core of a patent is the set of \textit{claims} which serve to define the scope of the legal protection granted by the patent.  Claims state, in technical and precise terms, the subject matter of the invention (or discovery), as well as its purpose, principal properties, the ways it operates and the methods it employs. The claims thus demarcate the technological territory controlled by inventors under the threat of suing for infringement. Claims have been a necessary part of U.S. patent applications since the enactment of the Patent Act of 1836. Although a patent is only required to include one claim, there is no upper limit on the number of claims that may be included in a patent. One claim must be identified as the ``controlling claim'' and this claim captures the most important aspect of inventive novelty in the patent. There is no specified relation between the number of claims and the number of technology codes used to classify a patent although one code, the ``original classification'' code does correspond to the controlling claim \cite{USPTO2012}. Every U.S. patent must have one and only one principal mandatory classification but may optionally include one or more additional ``discretionary" codes. Once the classification is completed a patent's codes fully encapsulates the aspect of novelty set forth in the claims \cite{StrumskyLoboLeeuw2012}. We clarify that the set of technology codes classifying any one invention is not a detailed listing of all of the technological functionalities utilized by the invention, but only of those functionalities pertinent to the invention's novelty.

A technology (or classification) code consists of two parts: a \emph{technology class} and a \emph{technology subclass}. Classes are major categories of patentable technology while subclasses delineate processes, structural features and functional specifications of the class. Subclasses have very detailed definitions and some subclasses are nested within hierarchical relationships to other subclasses. There are currently 474 technology classes and approximately 161,000 technology codes. A patent must have at least one code but there are no limits to how many codes may be assigned to a patent. As an example of a code, consider ``505/160'', which refers to the use of superconductor technology above a temperature of 30$^\circ$ Kelvin for measuring or testing mechanical, electrical, chemical, or physical properties.


\section{Invention as a Combinatorial Process}

One way to glean how important technological combination has been in the inventive process is to count how many patents are classified with a single technology code.  Seventy-seven percent of all patents granted between 1790 and 2010 are coded by a combination of at least two technology codes. Indeed, the combinatorial process has come to increasingly dominate inventive activity. But whereas in the 19$^{th}$ century nearly half of all patents are single-code inventions this proportion steadily decreased over the span of the 20$^{th}$ century, and currently stands at about 12\%. The mean number of codes, $\langle m \rangle$, classifying patents has accordingly been slowly increasing over the past two centuries indicating the steadily growing complexity of inventions.  Codes provide the alphabets for precise and parsimonious descriptions of technological capabilities; combining them into $m$ sets should be expected to result in descriptive words which are themselves precise to the point of uniqueness~\cite{StrumskyLobo2014}.

To better understand how technology codes and their combinations accumulate in the U.S. patent system, we provide the following stylized example that illustrates the difference between codes and their combinations, and clarifies the use of terminology. Suppose that at the start of a period, say year one, there are two patents, each of which is described by a set of technology codes (denoted by capital letters); in the following year three new patents are additionally granted which are also similarly described by a set of codes:

\begin{description}[labelindent=2cm]
    \item[$year=1$:] patent 1 = $\{A\}$, patent 2 = $\{A, B\}$
    \item[$year=2$:] patent 3 = $\{A, B\}$, patent 4 = $\{C, D,E\}$, patent 5 = $\{E\}$
\end{description}

\noindent The total inventions in the second year is simply the set of five patents $\mathcal{P} = \{ 1, 2, 3, 4, 5\}$, the collection of distinct technologies is identified by the set of technology codes $\mathcal{T} = \{A, B, C, D, \\E\}$, and the set of distinct combinations of codes used to describe inventions is $\mathcal{C}=\{A, AB, \\ CDE, E\}$. The individual technology codes $A$, $B$, $C$, $D$ and $E$ can be seen as individual words which together constitute a technological vocabulary. The U.S. Patent Office examiners draw from this vocabulary to describe the technological novelty embedded in an applied patent. Codes, much like words, can be used multiple times in different sentences, and some words can be used alone, as in the example above with $A$ and $E$. The cardinality of the three sets are $|\mathcal{P}|= 5$, $|\mathcal{T}|= 5$, and $|\mathcal{C}| = 4$.

We now apply the above formalism to the data on patents granted by the U.S. Patent Office to see how these variables can express the accumulation of inventions over 200 years. Figure 1(a) shows the time-series for patents $\mathcal{P}(t)$, technology codes $\mathcal{T}(t)$, and combinations of codes $\mathcal{C}(t)$ over the 1790 to 2010 period (following the convention in the patent research literature time is recorded as ``application year,'' the year a patent was successfully applied for). The growth exhibited by all three variables is clearly exponential for the first 80 years: straight lines in a semi-log plot.  During the first decades of the 19th century almost every invention brought to the attention of the Patent Office represented a new technology; the patenting system itself was an innovation and inventors rushed to turn a stock of existing technologies into patented inventions \cite{Khan2005}. A historically-minded reader may recognize the year 1870 as a marker for the period during which the U.S. became the world's dominant economy and the system of invention which historians have termed the ``American system of invention'' began to coalesce \cite{Rosenberg1982, Mokyr2002}.

Patents, codes and combinations accumulate in a similar manner up to about the year 1870, after which the increase in the number of technology codes slows down significantly while patents and combinations continue to grow in tandem.  The persistence of the combinatorial invention process can be seen more clearly in Fig. 1(b) where technology codes, $\mathcal{T}$, and combinations of codes, $\mathcal{C}$, are recorded against the number of accumulated inventions (the number of patents, $\mathcal{P}$). While invention introduces new codes at a much reduced rate (black circles), the introduction of new combinations proceeds unabated (blue triangles). In fact, the number of distinct combinations that have been used increases linearly with the number of patents, $\mathcal{C}(t) = \alpha \mathcal{P}$ with $\alpha \approx 0.6$:
\begin{equation}
  \Delta \mathcal{C} = 0.6 \Delta \mathcal{P}.
\end{equation}

\noindent An implication of this empirical relationship, solid red line in Fig. 1 (b), is that at 60\% of chance a new patent instantiates a new combination of technological functionalities while at complementary chance, 40\%, an invention uses exiting combinations of codes (thereby not contributing to an increase in $\mathcal{C}$). Figure 1(b) and equation (1) indicate that the process by which new technological combinations are introduced is systemic and persistent over almost two hundred years: the ratio $\Delta \mathcal{C}/\Delta \mathcal{P}$ is arguably $invariant$ for the entire period. The significance of the 0.6 coefficient can be better appreciated by considering two extreme scenarios. If the coefficient were equal to one, then any new invention would always be constituted by a new combination of codes; a very small value for the coefficient would, on the contrary, imply that the process of invention proceeds mostly by re-using existing combinations of codes (this would signify an invention process driven by improvements and refinements of existing inventions). The empirical record is found to be somewhere in between. That the fitted coefficient is slightly above half suggests that invention is proceeding mainly through new combinations of exiting technological capabilities. This systemic genesis of technological combinations contrasts sharply with the much reduced rate at which new technologies are introduced: this is the essence of the combinatorial process of inventions.

The de-linking between the generation of new technology codes and the growth of patents in Fig. 1 may indicate that the process of inventive combinatorics has enough components to sustain invention despite a slowdown in the introduction of new codes. Figure 2a plots the number of patented inventions, $\mathcal P$, as a function of accumulated technology codes, $\mathcal T$. From the moment when about 150,000 technological functionalities have been accumulated (late 19th century) the increase in the number of inventions proceeds with little addition to the existing stock of individual codes (as depicted by the near singularity in Fig. 2a). The conclusion again is that the process of invention is driven almost entirely by combining existing technologies.

The behavior shown in Fig. 2a raises a question regarding the sustainability of
the inventive process. How much further can combinatorial invention go given the finite size of new technology codes? Phrased differently, how big is the space of
technological possibilities and how much of this space has already been searched by existing
inventions? Since we know the number of codes $\mathcal{T}$ and the size of
combinations $m$ (shown in the inset in Fig. 2b) we can calculate a theoretical
bound for the possible number of combinations of codes and compare it with the empirical $\mathcal{C}$.  The number of possible combinations, $\mathcal{C}^{max}$, is $m$-combinations given a set of codes with $\mathcal{T}$ elements, expressed as
\begin{equation}
_\mathcal{T} \mathcal{C}_m = \frac{ \mathcal{T}!}{m! (\mathcal{T}-m)!}.
\end{equation}
To make Eq. (2) analytically tractable, we use Stirling's approximation for large $T$ and substitute $m$ with
a Gamma function, $\Gamma(m+1)$, that is, $\frac{(\mathcal{T}-m)^m}{e^m \Gamma(m+1)} (1-m/\mathcal{T})^{1/2-\mathcal{T}}$.  This approximation is convenient because $m$ is a variable distributed over a range of values with 1 as a lower bound, and we use average values up to the given year $t$, the $\langle m \rangle$, for simplicity; thus $m$ is not an integer anymore but a continuous variable.  The inset in Fig. \ref{fig:b} shows $\langle m \rangle $ over the years as well as $_\mathcal{T} \mathcal{C}_m$ and the actual $\mathcal{C}$.  We can immediately notice that the number of possible states is extremely sensitive to $m$: the value of the upper bound $\mathcal{C}^{max}$ with the maximum $m$ would be astronomically large compared to the one with the average $m$ (as shown in Fig 2b).

The occasional introduction of individual codes seems more than enough to sustain a combinatorial inventive search as Fig. 2 shows.  The huge gap between the possible and the actual number of combinations indicates that only a small subset of combinations become inventions and that re-combination can take place without much introduction of new technology codes once the set of codes is sufficiently large, which seems to be a general feature of combinatorial processes in nature, culture and technology \cite{Erwin2011, Wimsatt2013, Wagner2014}. As inventions have become more complex, here meaning that they are combining a greater number of codes, the number of combinatorial possibilities have likewise increased. It is the essential consequence of the combinatorial nature of invention that the size of the space of combinatorial possibilities has continued to increase despite the decrease in the rate at which new technology codes are introduced.

It is phenomenologically interesting that the gap increases over time despite the decrease in the entrance rate of new technological functionalities because of increasing $m$. The difference between the number of possible and realized technological combinations is reminiscent of a similar disparity in the biological domain: the number of realized genotypes is much less than would seem to be possible on the basis of recombination of genes~\cite{Carroll2001, Koonin2012}.  Likewise, the number of realized inventions is much less than would seem possible on the basis of available technologies available for combination. What accounts for this reduction in the ``phenotypic technological space''? The existence of ``technological trajectories,'' the restriction on technological change to certain developmental paths brought about by the institutionalisation of knowledge, skills sets and markets and professions, is possibly a candidate explanation \cite{Dosi1982}. So is the presence of technological ``path dependency'' \cite{David1985}. And of course many technologies will not be brought together as the resulting invention would not be of much use (ruling out inventions such as exploding prosthetics or espresso-making toothbrushes). It also seems plausible that topological features of the technology space, and intrinsic properties of the technologies themselves, would restrict which combinations are ever tried, but this is a question whose answer is beyond the scope of the research reported here.

As inventions have accumulated the size of the technological space has increased; how then has the space been searched and exploited? Equation (1) indicates that an invention either reuses a previously existing combination of technologies (at a rate of approximately 40\%) or introduces a new combination of technologies (approximately at a 60\% rate). Recall that patents can be granted to inventions that improve existing inventions thus reuse existing technological capabilities. The high rate of technological reuse is inconsistent with a random search of the space of technological possibilities. If one technological combination is randomly chosen from among all possible combinations, the probability of choosing the same combination twice is around $(10^4/10^6)^2 \approx 10^{-4}$ when $\mathcal T = 10,000$ in Fig 2 b. Furthermore, even if not all possible combinations are accessible but instead only a subset of them such that some combination have a higher probability of being selected twice, a random walk would generate a Poisson-like frequency distribution for the reuse of combinations. Figure 3 shows the frequency distribution for the use of (a) individual technologies (represented by technology codes), and (b) combinations of technologies (represented by combinations of technology codes): both quantities show somewhat heavy-tail distributions which conflicts with a model of technological search as a random walk.

The empirically identified distributions of frequency use up to the year 2013 (Fig. 3) tell us that some individual technology codes and combinations of codes are used quite intensely while others are hardly used at all. This heterogeneous frequency distribution is known to be a key characteristic of the Yule-Simon process, or ``urn process'', in which balls are added to a growing number of urns with a probability linear to the number of balls already in the urn (The urn process is often used as a model to explain the distribution of biological taxa and subtaxa \cite{Simon1955}.) In the Yule-Simon process a skewed frequency distribution results from the self-reinforcing property sometimes referred to as ``the rich get richer''. This effect would manifest itself in a positive correlation between the number of times combinations of technology codes have been used and their vintage (that is, how long they have been available in the inventory of technologies, which may have increase chances to be used). Figure 4 shows the frequency of reuse for codes, in technology combinations, after they first appeared in the patent record. Although there is time-lag effect, a ``rich get richer'', or ``first mover advantage'', is hardly evident in the plots.  Just because a certain technology has been around for longer does not particularly increase the likelihood that it will be used more times.  This is in consistent with the observed dynamics of patent citations exhibiting an aging term weighting the likelihood of citing previous patents ~\cite{Valverde2007}.

\section{Invention: Broad or Narrow Combinations?}

The combinatorial process generating inventions proceeds via a preponderance of \textit{novel} combinations of technologies.  But another dimension of inventive novelty is revealed by examining how disparate individual codes combined by an invention.  Are the technological combinations brought together in a patented invention closely related or distinctly different ones? Intuitively, an invention which integrates technological functionalities drawn from ``distant'' domains should be considered more novel than an invention whose constituent functionalities represent variations on one technological theme.

We can categorize technological combinations as ``narrow'' or ``broad'', in effect operationalize a notion of ``technological distance", by relying on the basic feature of the USPTO classification scheme. Recall that classes denote major boundaries between technologies, while subclasses serve to specify processes, features, and functionalities. Although different codes denote distinct technological capabilities, codes sharing a technology class are in closer technological proximity than codes drawn from different classes. ``Narrow'' here means that the technology codes used to classify the novelty of an invention are similar to each other (i.e., are based on the same technology class), as opposed to ``broad'' meaning that the technology codes represent different classes. Even a patented invention which is undeniably considered a ``breakthrough invention'' -- patent \#4,237,224 for the recombinant DNA technique -- is described as bringing together 24 distinct technologies of which 20 are drawn from the same class.

The notions of ``broad" and ``narrow" technological combinations provide us with one way to assess just how novel has patenting activity been: measure the proportion of patents granted over a year's time whose classification involves the use of multiple codes which are all based in the same technology class. Figure 5 plots, over a 200 year span of inventive activity, the of percentage of multiple-codes patents which are ``narrow". The percentage hovers around 44\% although the time-series clearly displays two distinct regimes: in the decades before 1930 about half of the technological combinations were ``narrow", a proportion which greatly decreased, to about 30\%, in the decades following WWII. (This pattern is consistent with the often-made observation that the post-WWII period was a very inventive and innovative period for the U.S. economy ~\cite{Gordon2012}.) But starting around 1970 the proportion of technological combinations (that is, inventions) which are ``narrow" began to increase and currently stands at about 50\%. The heightened importance of ``narrow" inventions coincided with the dramatic increase in the rate of patenting as over 50\% of all patents ever granted by the USPTO have been granted since 1980. It may be very hard to sustain invention solely on the basis of truly novel technological combinations.

\section{Discussion}

Arthur and Polak \cite{ArthurPolak2006} eloquently state the combinatorial view of technological change: ``New technologies are never created from nothing. They are constructed -- put together -- from components that previously exist; and in turn these new technologies offer themselves as possible components -- building
blocks -- for the construction of further new technologies.'' (p.23) By using patent technology codes to identify distinct technologies and their combinations we are able to systematically and empirically study the combinatorics of invention. We find that the combination of technologies has indeed been the major driver of invention, reflected in an invariant introduction of new combinations of technologies engendered by patented inventions. The introduction of new technological functionalities plays a minimal role in fueling invention once the system is mature. Instead, tinkering, gradual modification and refinements are very important in pushing invention forward.
We also find that a few technologies and their combinations are utilized much more intensely in the construction of inventions, and that the level of utilization is not correlated with how long the technologies have been available in the system.

By definition all patented inventions are ``novel", but not all novelty is created equally. U.S. patent law allows for patents to be granted to inventions which represent improvements over existing inventions. This implies that differentiated levels of novelty are inherent in the patenting system. Utilizing patent technology codes to characterize the combinatorial process of invention makes it possible to assess the novelty of inventions on the basis of the technologies combined in inventions (in contrast to other patents cited as part of prior art) \cite{Valverde2007}. The novelty instantiated by patented inventions stems from combining technologies -- not from the utilization of new technologies per se -- with a slight preponderance of the technological combinations constituting inventions being new having not been seen before in the patent record.  And combinatorial invention brings together related technologies as frequently as technologies from distinct domains. These observations -- on the balance between old and new technological combinations -- are broadly consistent with recent work on scientific papers which finds that high-impact (and presumably high-quality) papers combine atypical and conventional knowledge \cite{Uzzietal2013}.

The language of ``exploration'' and ``exploitation" introduced by March \cite{March1991} to describe organizational learning as a search process comes to mind when thinking about inventive search.  Novel combinations and interdisciplinary combinations (involving distinct technological domains) signify inventors engaging in \textit{exploration} while the reuse of combinations and the integration of similar technologies signal inventors engaging in less risky search, i.e., \textit{exploitation}. The search strategy implicit in the process of invention seems to be poised halfway between \textit{exploration} and \textit{exploitation} in what might be a nearly optimal search strategy.


There has been a body of literature to understand technlogical evolution through mechanics of Darwinian evolution \cite{Wagner2014}.  Our results highlight tantalizing, and empirically grounded, similarities with another generative combinatorial search process, biological evolution. Firstly, only a relatively small number of information building blocks -- protein-coding genes -- have been involved in the construction of most genomes \cite{Erwin_etal2011}. Secondly, biological evolution is a historical path-dependent process in which the order and success of adaptations depends upon the order in which they occur \cite{Stern2010}. Lastly, the exploration of possible morphologies through morphological space has not been random but has instead been guided by selection \cite{Thomasf_etal1993}.  Studying patent, comparative and systemic records of inventions, will open a way to make quantitative assessments for a counterpart of these features of biological evolution in technological evolution.

\section{Acknowledgements}
Youn would like to acknowledge the support of research grants from the National Science Foundation (no. SMA-1312294) and the James S. McDonnell Foundation (no. 220020195); Lobo and Strumsky acknowledge the support of research grants from the National Science Foundation and Department of Energy. The authors thank Aaron Clauset and Doyne Farmer for helpful comments on an earlier version of the paper.

\begin{figure}
	\centering
	\includegraphics[width=4in]{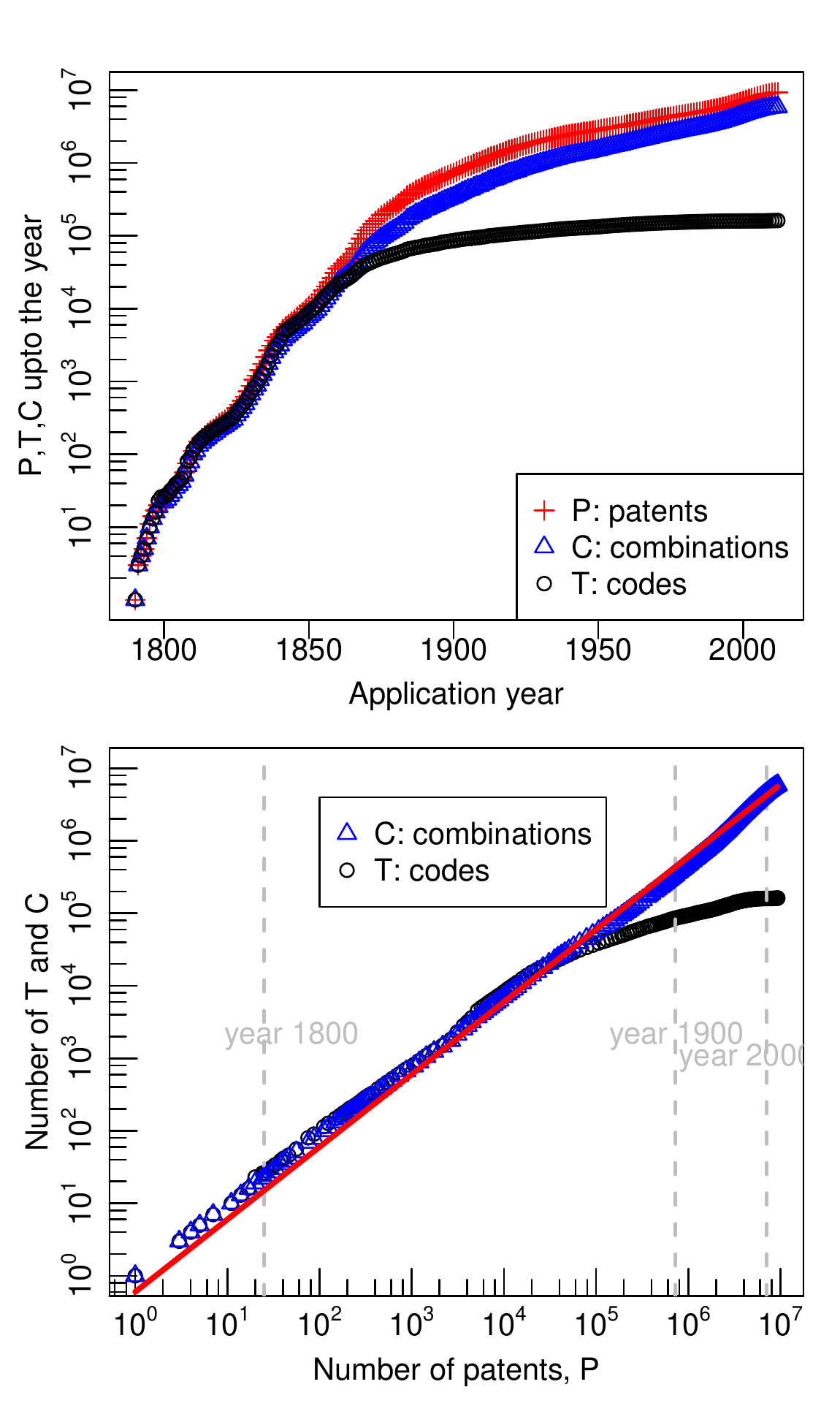}
	\caption{
The number of patents, technology codes, and combinations of codes as inventions
accumulate in the system. The panel (a) shows their increase with year, and the
panel (b) shows them as implicit functions of time by expressing
one of their variables, the number of patents.  The gray dashes in (b) mark the
number of patents corresponding the year.
Because the number of patents increase exponentially in time, the gaps between year
marks get shorter and shorter. The black line is a linear fit with combinations $C$.
  \label{fig:a}
}
\end{figure}


\begin{figure}
	\centering
	\includegraphics[width=4in]{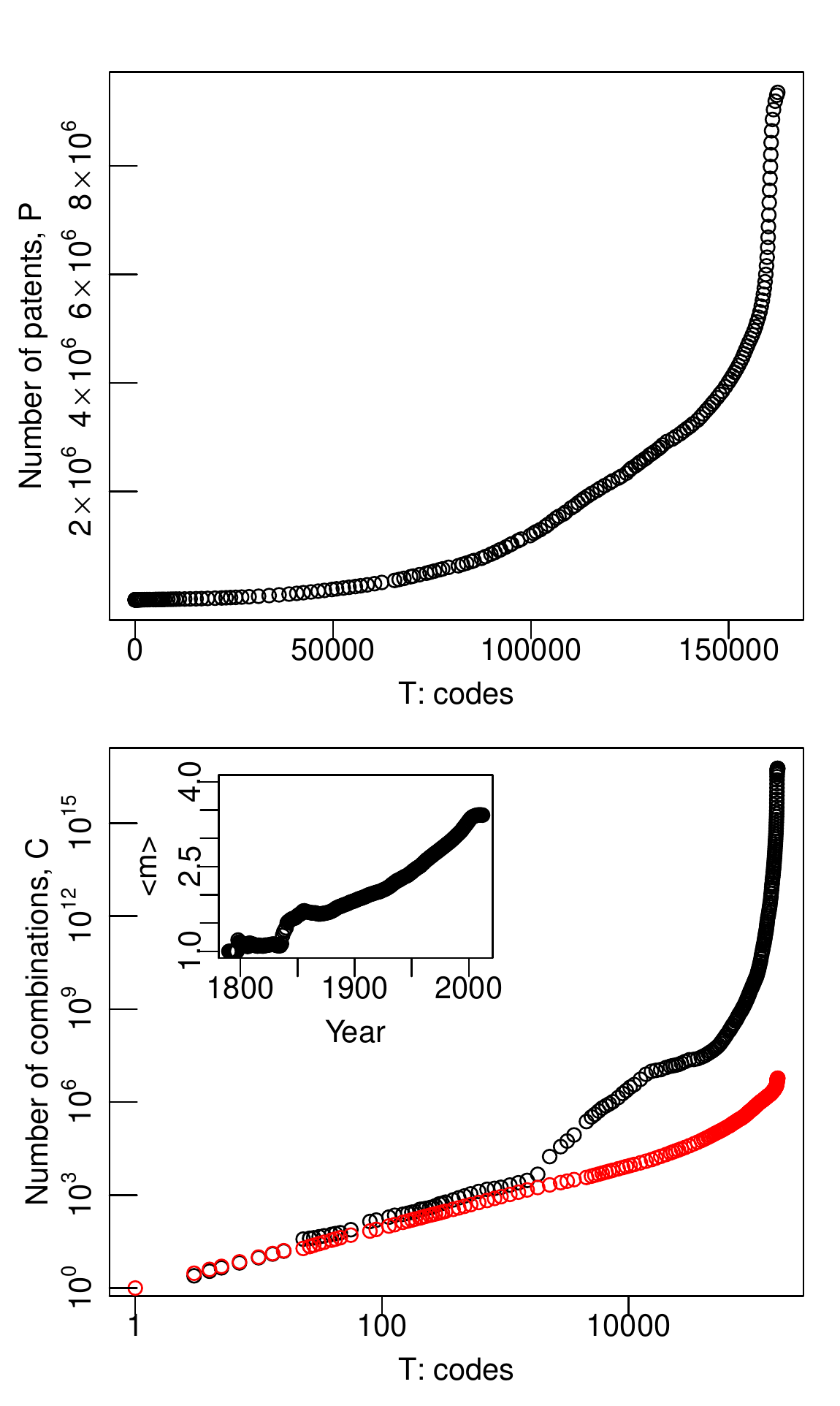} 
	\caption{  The number of patents $P$ as the number of codes accumulate in the system (a) and the theoretical bound for possible $m$-combinations given a set of codes of
	$T$ elements, that is, $_T C_m$, as is denoted in black circles (b). $T$ and $m$ are drawn
	from the empirical data. The red circles are the number of combinations that
	have actually been used for inventions. The inset shows the empirical $m$ over time.
  \label{fig:b}
}
\end{figure}

\begin{figure}
	\centering
	\includegraphics[width=5in]{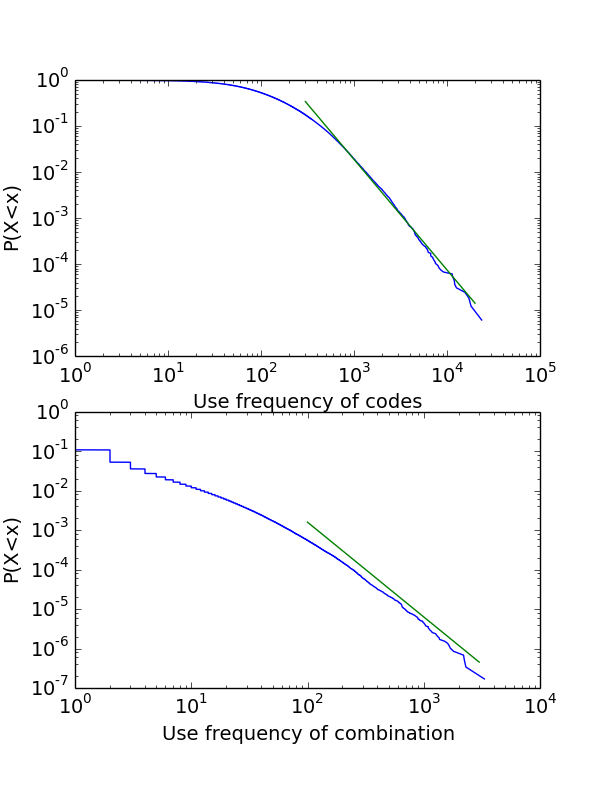} 
	\caption{ Cumulative frequency distributions of (a) use of codes and (b) use of combinations
  denoted as circles, $P(X>x)$, where $x$ is use frequency until the year 2013. The solid lines guide $x^{-2.4}$ for both tails.
  \label{fig:c}
}
\end{figure}

\begin{figure}
	\centering
	\includegraphics[width=5in]{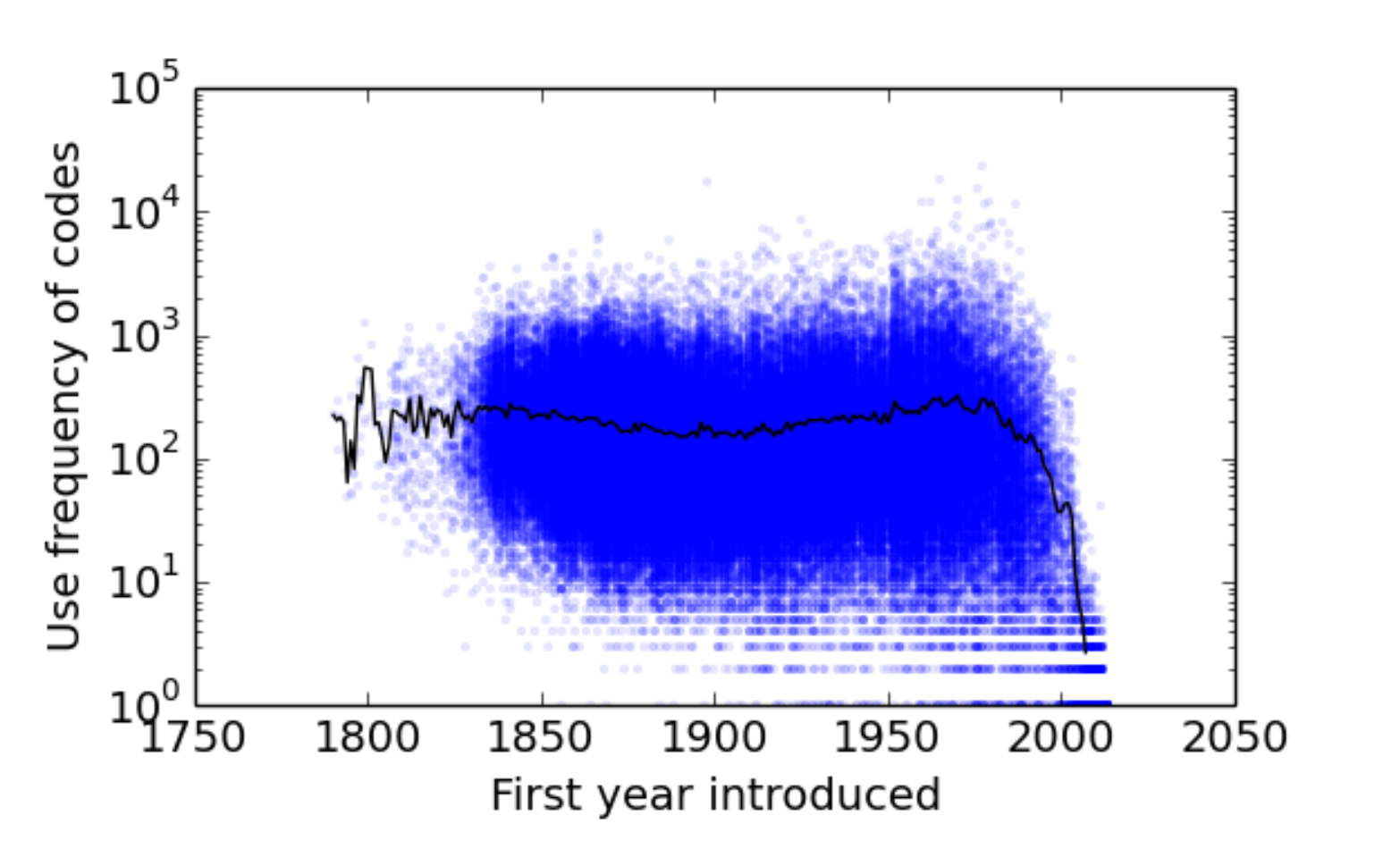}
	\caption{  Codes were plotted with the years when they were first applied, and corresponding frequencies of their recurrences. Average of each year is denoted as solid line. There is no systemic correlation observed: no advantage of being the first mover. The drop found is attributed to both the overdued patents not yet granted, and the time lag for inventors to utilize building blocks
	\label{fig:d}
}
\end{figure}

\begin{figure}
	\centering
	\includegraphics[width=5in]{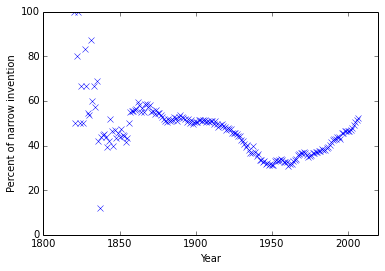}
	\caption{
The percent of patents that bring a set of codes within one class (narrow invention) per each year.  Notice the decrease of narrow inventions (the increase of broad inventions) during post World War II.
	\label{fig:e}
}
\end{figure}

\clearpage
\baselineskip 24pt
\bibliographystyle{Science}

\end{document}